\documentclass[12pt,preprint]{aastex}

\shorttitle{Evidence for polar jets as precursors of polar plume formation}
\shortauthors{Raouafi et al. 2008}

\begin{document}


\title{Evidence for polar jets as precursors of polar plume formation}

\author{N.-E. Raouafi\altaffilmark{1}, G. J. D. Petrie\altaffilmark{1}, A. A. Norton\altaffilmark{1},
C. J. Henney\altaffilmark{1}, and S. K. Solanki\altaffilmark{2}}


\altaffiltext{1}{National Solar Observatory, Tucson, AZ 85719, USA. nraouafi@nso.edu;
petrie@nso.edu; norton@nso.edu; henney@nso.edu.}
\altaffiltext{2}{Max Planck Institute for Solar System Research, 37191 Katlenburg-Lindau, Germany.
solanki@mps.mpg.de.}

%%%%\altaffiltext{2}{Society of Fellows, Harvard University.}
%%%%\altaffiltext{3}{present address: Center for Astrophysics,
%%%%    60 Garden Street, Cambridge, MA 02138}
%%%%\altaffiltext{4}{Visiting Programmer, Space Telescope Science Institute}
%%%%\altaffiltext{5}{Patron, Alonso's Bar and Grill}


\begin{abstract}

Observations from the Hinode/XRT telescope and STEREO/SECCHI/EUVI are utilized to study polar
coronal jets and plumes. The study focuses on the temporal evolution of both structures and their
relationship. The data sample, spanning April 7-8 2007, shows that over $90\%$ of the 28 observed
jet events are associated with polar plumes. EUV images (STEREO/SECCHI) show plume haze rising from
the location of approximately $70\%$ of the polar X-ray (Hinode/XRT) and EUV jets, with the plume
haze appearing minutes to hours after the jet was observed. The remaining jets occurred in areas
where plume material previously existed causing a brightness enhancement of the latter after the
jet event. Short-lived, jet-like events and small transient bright points are seen (one at a time)
at different locations within the base of pre-existing long-lived plumes. X-ray images also show
instances (at least two events) of collimated-thin jets rapidly evolving into significantly wider
plume-like structures that are followed by the delayed appearance of plume haze in the EUV. These
observations provide evidence that X-ray jets are precursors of polar plumes, and in some cases
cause brightenings of plumes. Possible mechanisms to explain the observed jet and plume
relationship are discussed.

\end{abstract}





\section{Introduction}

Recent space missions, such as Hinode \citep{Kosugi07} and STEREO \citep{Kaiser08}, and
ground-based facilities such as SOLIS \citep{Keller03} provide a set of data unprecedented in
quality and cadence. The complementary observations from the different instruments provide the
necessary spatial, temporal and temperature coverage to observe the dynamics of jets and polar
plumes, helping to form a more complete picture of these structures.

X-ray jets occur almost everywhere in the solar corona \citep[see][]{Shibata92}, in particular
in the polar holes. They are characterized by their transient nature and often appear as
collimated high-temperature emissive beam guided by open magnetic flux \cite[length of $10^5-10^6$
km and collimated widths of $\sim10^4$ km; see][]{Cirtain07}. \cite{Cirtain07} reported that the
plasma outflow speeds within X-ray jets range from $\sim100$ to $\sim1000$ km~s$^{-1}$ and that
Alfv\'en waves are responsible for the high outflow velocities.

In contrast, polar coronal plumes are observed to be hazy in nature without sharp edges, as seen in
extreme ultra-violet (EUV) images from SOHO/EIT \citep{Boudin95} and STEREO/SECCHI/EUVI
\citep{Howard08}. Plumes are also observed to be significantly wider than X-ray jets
\cite[$\sim20-40$~Mm; see][]{Wilhelm06} and reach several solar radii in height
\cite[see][]{DeForest97}. Plumes are brighter, cooler and the plasma outflows are smaller than in
inter-plumes \citep[see][]{DeForest97,Wilhelm98,Raouafi07}.

%%Plumes are brighter than the background due to higher densities compared
%%to inter-plume regions. Plumes are cooler and the plasma outflows are smaller than in inter-plumes
%%\citep[see][]{DeForest97,Wilhelm98,Raouafi07}.

\begin{figure*}
\begin{center}
\includegraphics[width=0.5\textwidth]{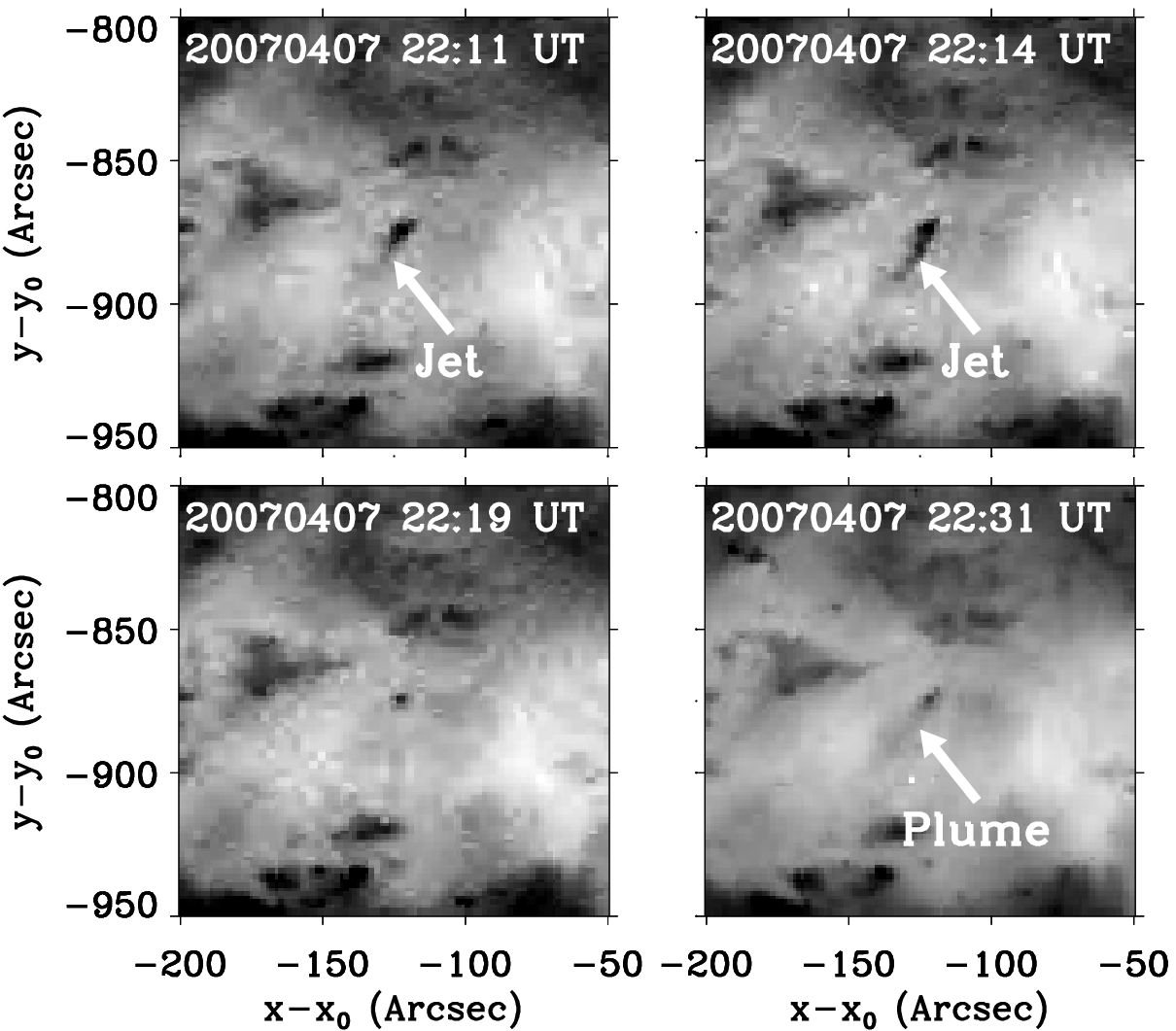}
\caption{STEREO/SECCHI/EUVI ``A'' images illustrating an EUV polar jet evolving into a polar plume
with a time delay of 10-15 minutes. \label{EUV_fig_jet_plume_070407_2200}}
\end{center}
\end{figure*}

Recent studies of jets and polar plumes \citep[X-ray and EUV; see][]{Wang98,MInsertis08} treat
these coronal structures independently and the relationship between them is not investigated. The
present research is motivated by the fact that polar X-ray and EUV jets and plumes usually share
common properties. Both are episodic in nature and occur at magnetic field concentrations that
coincide with the chromospheric network where both structures form through flux emergence
\citep[see][]{Canfield96,Wang98}. Studying the relationship between jets and plumes is important to
understand their formation processes, evolution and the eventual contributions to the solar wind
and heating of the plasma in the polar coronal holes. The present work is motivated by the
observations of polar jets evolving into plumes such as the one shown in
Fig.~\ref{EUV_fig_jet_plume_070407_2200}. The aim of the paper is to investigate the relationship
between these prominent coronal structures.

\section{Observations and Data Analysis}

The XRT telescope \citep{Golub07} on Hinode provides high resolution images ($\approx1-2\arcsec$
depending on the location within the field of view) of the solar corona at temperatures ranging
from 1-20~MK. Observations of the southern coronal hole from XRT were utilized to study the
evolution of polar X-ray jets and their relation with plumes. The data cover several time intervals
on April 7-8, 2007 (07 April: 03:30 - 06:59 UT and 18:29 - 23:59 UT; 08 April: 11:49 - 17:59 UT and
21:30 - 22:59 UT) with a cadence of less than a minute. The data were corrected for instrumental
effects utilizing XRT-calibration procedures.

A total of 28 X-ray jets were identified, with at least two recurring events within an hour. Most
of the events are characterized by sharp collimated beams. The observed jets have different
properties with regards to brightness, spatial extension, lifetime, and evolution. The bright point
at the base of each jet is enhanced in brightness with every eruption and then fades after the jet
is no longer observed. 

In addition, 171~{\AA} images from the STEREO/SECCHI satellite ``A'' were utilized to study EUV
features in relation to the identified X-ray events. Particular attention was given to the presence
of plume material during or after the eruption of jets. The choice of 171~{\AA} was dictated by the
adequate temperature corresponding to polar plume emissions.

%%Images recorded every one and a half minutes were selected for the present study.

\section{Results}

\begin{figure*}
\begin{center}
\includegraphics[width=\textwidth]{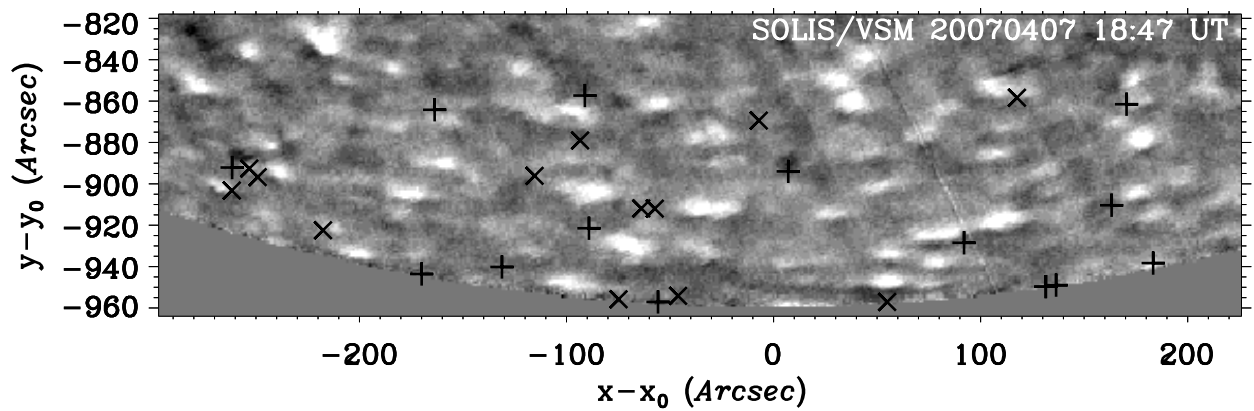}
\caption{SOLIS/VSM line-of-sight chromospheric magnetogram (Ca~{\sc{ii}} 8542~{\AA}) showing the
location of the X-ray jets observed by Hinode/XRT on 2007 April 7$^{\rm{th}}$ (`+' signs) and
8$^{\rm{th}}$ (`$\times$' signs). Their displacement due to solar differential rotation is taken
into account. \label{solis_070407_jet_plume}}
\end{center}
\end{figure*}

Fig.~\ref{solis_070407_jet_plume} displays the LOS-chromospheric magnetogram (Ca~{\sc{ii}}
8542~{\AA}) of the south pole on April 7, 2007 recorded by the SOLIS/VSM instrument
\citep{Henney08}. Spatial location of the X-ray jets on April 7$^{\rm{th}}$ and 8$^{\rm{th}}$ are
marked by `+' and `$\times$' signs, respectively. No SOLIS/VSM chromospheric magnetograms were
available for April 8, 2007. The solar rotation effect on the events' spatial locations has been
corrected using the model by \citet{Howard90}. It is clear that most jet events, in particular
those of April 7$^{\rm{th}}$, are rooted in or near magnetic flux concentrations. At the base of
bright jets are relatively large flux elements of one polarity surrounded by more diffuse flux of
the opposite polarity (see Figs.~\ref{solis_070407_jet_plume} and \ref{fig_letter_080407_0332UT}).
Weaker and short-lived jets are based in areas of more diffuse magnetic flux.

Top panels of Figs.~\ref{fig_letter_080407_0332UT}-\ref{fig_letter_080408_2240UT} show a sample of
nine X-ray jets recorded by Hinode/XRT on April 7$^{\rm{th}}$ and 8$^{\rm{th}}$ 2007, respectively.
The different events are indexed xj$_i$ ($i=1-9$) according to the time of their appearance.
Although the brevity of the polar observation sequences did not allow us to determine the real
lifetime of several events, jet lifetimes are estimated to range from minutes to a few tens of
minutes with a number of events recurring within an hour, such as the event xj$_1$.

The middle and bottom panels of Figs.~\ref{fig_letter_080407_0332UT}-\ref{fig_letter_080408_2240UT}
display EUV images of the southern polar region corresponding to the X-ray observations. The data
cover time intervals spreading over several hours after the disappearance of the X-ray events. A
number of X-ray jet events are also present in EUV images (i.e., xj$_1$ and xj$_2$ in
Fig.~\ref{fig_letter_080407_0332UT}a and corresponding EUV structure in
Fig.~\ref{fig_letter_080407_0332UT}d; similarly xj$_7$ in Fig.~\ref{fig_letter_080408_2240UT}a and
Fig.~\ref{fig_letter_080408_2240UT}g-h). Some of these events look brighter and sharper in EUV than
in X-ray (see xj$_2$ in Fig.~\ref{fig_letter_080407_0332UT}a and EUV counterpart in
Fig.~\ref{fig_letter_080407_0332UT}d), perhaps for plasma temperature reasons. This highlights that
X-ray and EUV jet events are contiguous when plasma conditions allow emission in both temperature
ranges. 

\begin{deluxetable}{ccc}
\tabletypesize{\scriptsize}
\tablecaption{Correlation of X-ray jets to the EUV jet plume events shown in
Figs.\ref{fig_letter_080407_0332UT}-\ref{fig_letter_080408_2240UT}.
Events close to the limb (i.e., xj$_3$, xj$_4$ and xj$_8$) are not listed.\label{corr_jet_plume}}
\tablewidth{0pt}
\tablehead{
\colhead{X-ray jet} & \colhead{EUV jet} & \colhead{Polar plume} }
\startdata
xj$_1$ (Fig.\ref{fig_letter_080407_0332UT}a) & Fig.\ref{fig_letter_080407_0332UT}d &
Fig.\ref{fig_letter_080407_0332UT}e-i    \\
xj$_2$ (Fig.\ref{fig_letter_080407_0332UT}a) & Fig.\ref{fig_letter_080407_0332UT}d &
Fig.\ref{fig_letter_080407_0332UT}e-f    \\
xj$_5$ (Fig.\ref{fig_letter_080407_0332UT}c) & -- & Fig.\ref{fig_letter_080407_0332UT}g    \\
xj$_6$ (Fig.\ref{fig_letter_080408_2240UT}a-c) & Fig.\ref{fig_letter_080408_2240UT}d &
Fig.\ref{fig_letter_080408_2240UT}e-i     \\
xj$_7$ (Fig.\ref{fig_letter_080408_2240UT}a) & -- & Fig.\ref{fig_letter_080408_2240UT}g-i    \\
xj$_9$ (Fig.\ref{fig_letter_080408_2240UT}c) & -- & Fig.\ref{fig_letter_080408_2240UT}g-i
\enddata
\end{deluxetable}

The EUV data show that a significant number of polar jet eruptions are followed by rising polar
plume haze with a time delay ranging from minutes to hours. Table~\ref{corr_jet_plume} summarizes
the correlation and corresponding figures between the different X-ray and EUV events. A good
example of plume haze appearing after a jet is given by the event xj$_6$, where collimated plasma
emission is observed both in X-ray and in EUV images (see Fig.~\ref{fig_letter_080408_2240UT}). The
xj$_6$ event first appeared in X-ray images earlier than 21:31 UT (no X-ray data available to
determine the exact start time). This event dimmed around 21:47 UT and reappeared again around
21:58 UT. The collimated EUV emission lasted longer than the X-ray one and evolved gradually into a
wider and hazy structure that lasted for several hours, showing a polar plume with time-varying
emission. Events xj$_3$, xj$_4$ and xj$_8$ were adjacent to off-limb plume emission locations.
Cases of polar jets erupting within the base of ongoing plumes resulted in emission enhancement of
the latter (compare P$_{07}$ in Fig.~\ref{fig_letter_080407_0332UT}f \& h and
Fig.~\ref{fig_letter_080407_0332UT}i; and P$_{08}$  in Fig.~\ref{fig_letter_080408_2240UT}d-i).

\section{Discussion}

\begin{figure*}
\begin{center}
\includegraphics[width=\textwidth]{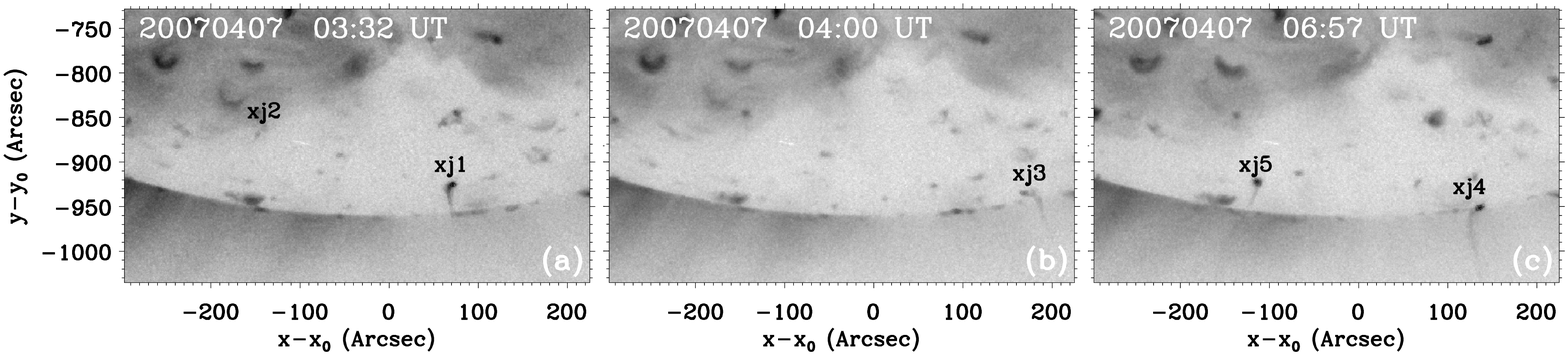}
\includegraphics[width=\textwidth]{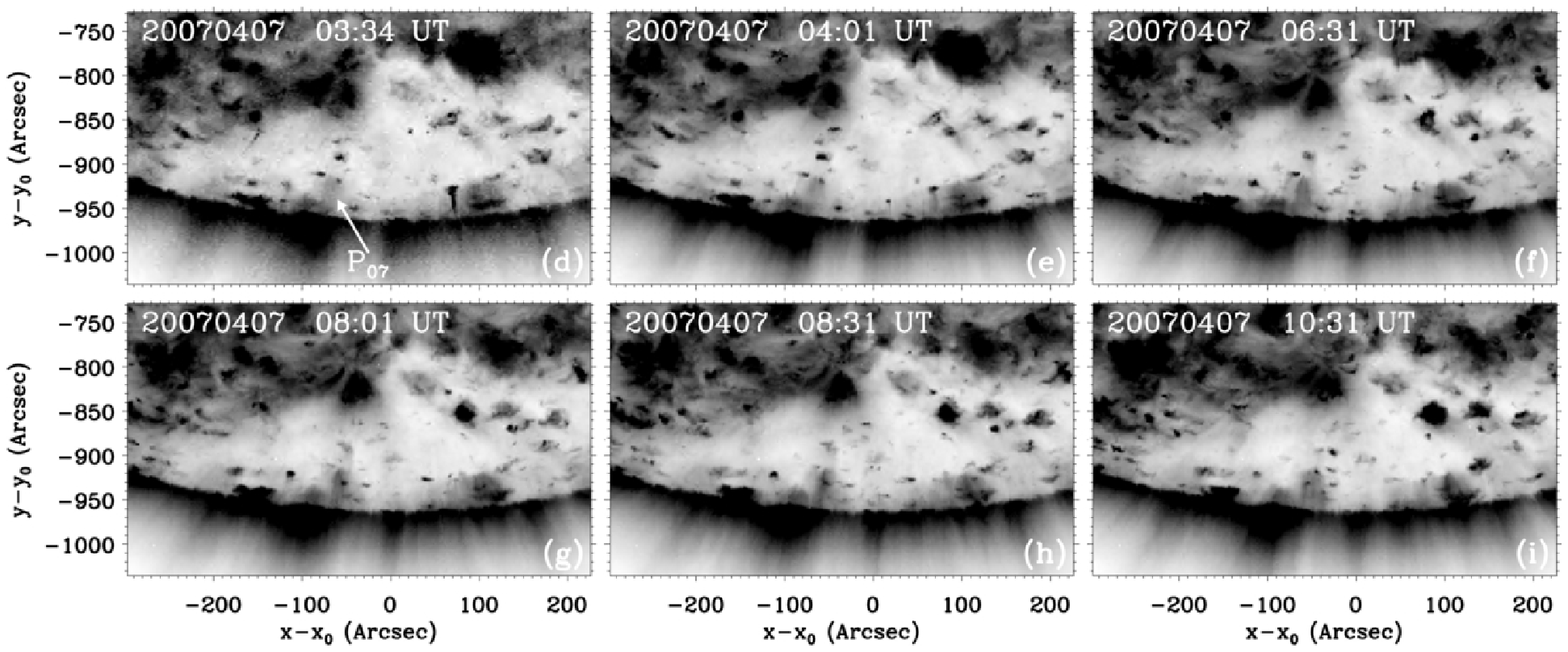}
\caption{Top: Hinode/XRT snapshots of the southern polar coronal hole recorded on 07 April 2007
showing several X-ray jet events. X-ray jets are labeled by xj$_i$ $(i=1,...,5)$ according to their
time of appearance. Middle and bottom: 171~{\AA} images from STEREO/SECCHI/EUVI ``A'' of the
southern polar hole of the same day. Polar plume haze clearly rise from the same locations as X-ray
and EUV jets with a time delay ranging from minutes to hours. A number of short-lived,
jet-like events also occur at the base of polar plumes. 
\label{fig_letter_080407_0332UT}}
\end{center}
\end{figure*}

X-ray and EUV observations indicate that more than 90\% of the jets observed in the southern polar
hole on April 7-8, 2007 are associated with plume haze. 70\% of these jets are followed by polar
plumes with a time delay ranging from minutes to tens of minutes. Emission of pre-existing plumes
is enhanced after every jet eruption within their base. A number of prominent plumes (e.g.,
P$_{07}$ and P$_{08}$) show evidence for short lived, jet-like events in the EUV that occur within
the plume base (see the sharp structures Fig.~\ref{fig_letter_080407_0332UT}(f) \& (h) and the
several bright points in panel (i)). Jet-like events ensure the continuous rise of haze and may
contribute to the change in plume brightness \cite[see][]{DeForest97}.

%\citet{Wang98} points out that polar plumes lag the formation of bright points by several hours. 

\begin{figure*}
\begin{center}
\includegraphics[width=\textwidth]{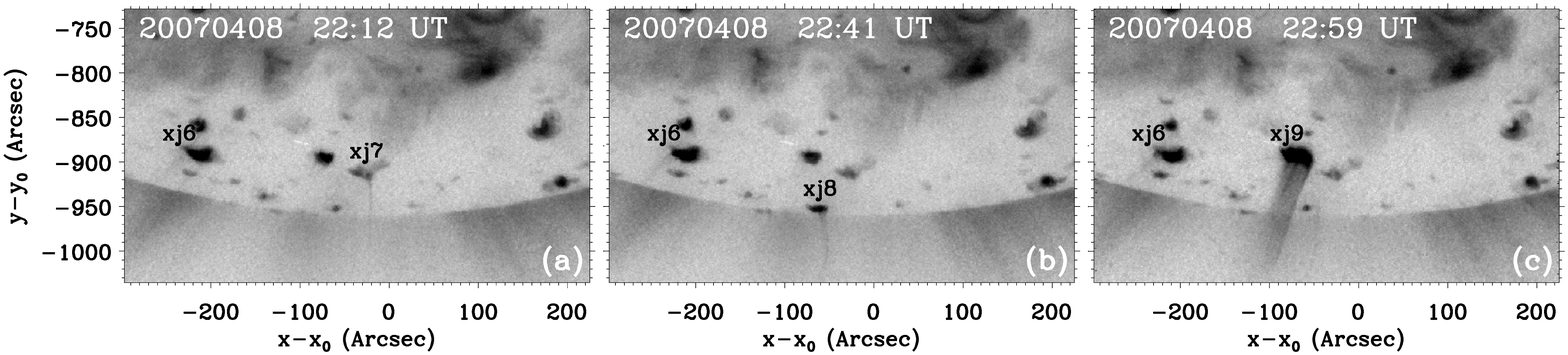}
\includegraphics[width=\textwidth]{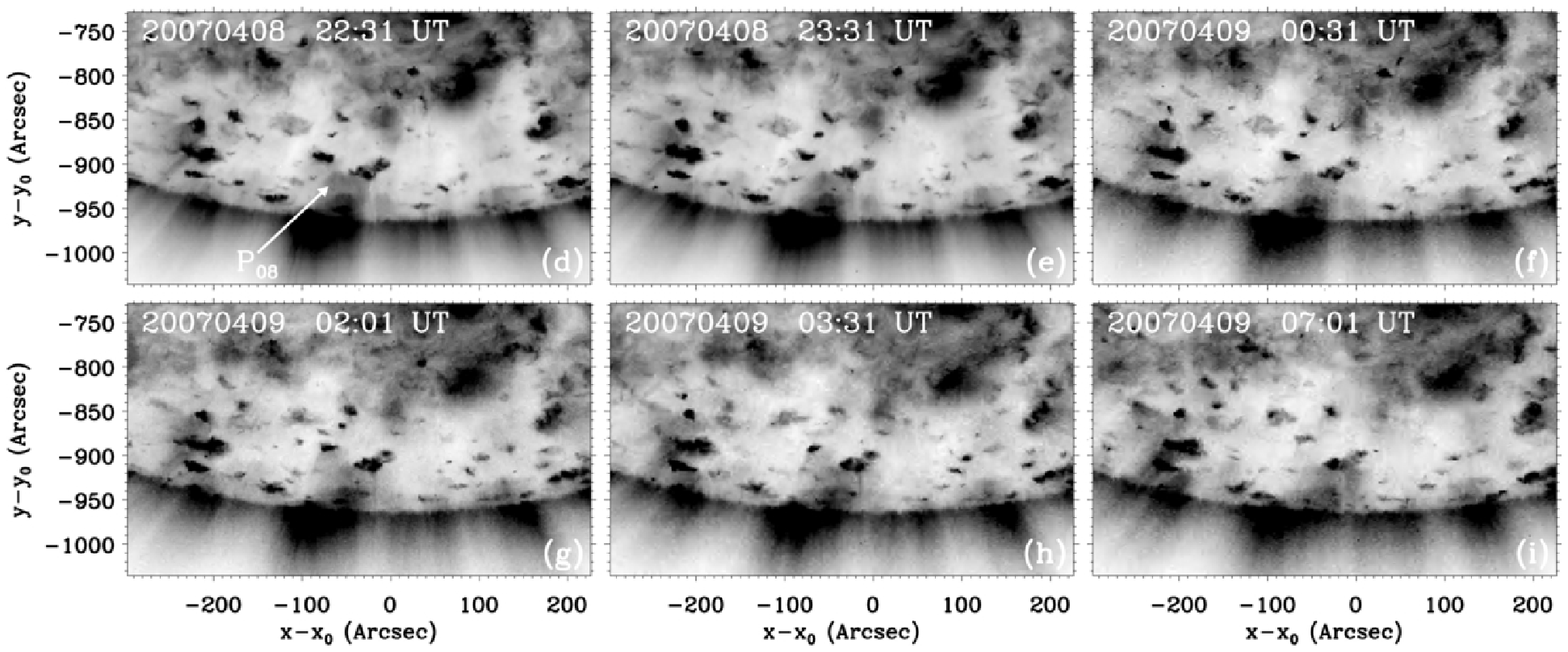}
\caption{Same as Fig.~\ref{fig_letter_080407_0332UT} but for 08 April 2007. EUV images are
recorded on 08 and 09 April 2007. \label{fig_letter_080408_2240UT}}
\end{center}
\end{figure*}

The event xj$_7$ in Fig.~\ref{fig_letter_080408_2240UT} is an interesting case. It was observed in
X-rays from 21:58 - 22:16 UT on April 8, 2007. Fig.~\ref{fig_letter_080408_2240UT}(d-e) shows an
EUV collimated structure similar to the one observed in X-rays more than three hours earlier. This
may be caused by the plasma being heated to several MK and then becoming visible in X-rays, then
gradually cooling down until it appears in the EUV range. More data needs to be analyzed to confirm
the plausibility of this hypothesis.

The event xj$_9$, illustrated by Fig.~\ref{fig_letter_080408_2240UT}(c), is also peculiar and
lasted less than 30 minutes. A narrow, collimated beam of plasma rose from the left edge of the
large bright point with a shape typical of X-ray jets. It evolved rapidly and after 4-5 minutes the
base width of the emission began to widen to cover the whole bright point. The width of the
emitting structure exceeded 20~Mm, which is the typical width of polar plumes
\cite[see][]{Wilhelm06}. EUV images showed a faint haze several hours after the X-ray event (see
Fig.~\ref{fig_letter_080408_2240UT}d-i). GONG magnetograms show that the flux at the base of xj$_9$
weakened during the event's lifetime. We believe that the initial jet event evolved into a plume
due to significant emerging magnetic flux causing a catastrophic magnetic reconnection on a
relatively short time scale but over a large spatial area. This may allow dissipation of the
magnetic energy budget of the structure over a short period of time with an associated ejection of
a significant amount of material over a relative large spatial scale, unlike other jet-plume events
that develop over intervals of several hours. This type of event is recorded twice in the data set
utilized here.

It is likely that jets play a key role in the formation process of polar plumes. Both coronal
structures share numerous common characteristics, i.e., a magnetic field of mixed polarities at the
base, leading to magnetic reconnection. We believe that the magnetic flux emergence causes the jet,
opening of previously closed flux results in plume. Jet eruption seems to be the result of
gradually emerging magnetic flux from the solar interior that suddenly reconnects on a small scale
with the ambient photospheric field, leading to a collimated beam of plasma rising in the corona
\citep[e.g.,][]{Yokoyama95}. EUV images show that coronal plume haze is observed following the jet
events. They also provide evidence for several small bright points and short-lived, jet-like events
within the base of the plume. These may be the results of magnetic reconnection at smaller
spatio-temporal scales that modulate and sporadically brighten pre-existing polar plumes. This is
most often seen in long-lived polar plumes, since several phases of reconnection can develop in a
single long-lived structure. However, fast opening of magnetic flux can allow a plume to develop
almost immediately such as in the case of the xj$_9$ event.

The transition from fast, impulsive, magnetically-driven dynamics of reconnection to the thermal
expansion of newly liberated gas along opened magnetic field could explain the time delay observed
between the jet and plume events. On the one hand, the jet eruption is the result of fast and
explosive dissipation of magnetic energy on a short time scale. On the other hand, the plume might
be a result of a pressure gradient within the open flux, which would lift the plume material in the
corona. This hypothesis is supported by the fact that plasma outflow velocities in plumes are
measured to be rather low up to $\sim1~R_\sun$ above the solar surface. The continuous emergence of
magnetic flux at a slow rate and relatively large scale might ultimately create a sizable bundle of
newly opened flux, allowing in turn a significant plume of escaping plasma to develop.

It is beyond us to simulate the development of a jet into a plume in an MHD model.  However, some
basic physics of such a development can be anticipated.

If a bipolar field emerges into a unipolar, open field region, then the two fields are not, in
general, exactly parallel across the boundary between them.  Then, according to Parker's (1994)
theory, a magnetic tangential discontinuity forms and current dissipation and field reconnection
become inevitable at this boundary. Any two non-parallel fields can be resolved into parallel and
anti-parallel components. The anti-parallel components will mutually annihilate at the
discontinuity. The dissipated magnetic energy is partially converted to kinetic and thermal energy,
which would cause a jet of energized plasma to escape along the open field next to the dissipating
current sheet. Whenever some quantity of open flux is locally annihilated along the current sheet
an equal quantity of closed flux must become open for magnetic flux continuity
(${\bf\nabla}\cdot{\bf B}=0$).  This open flux can allow a plume of thermally expanding plasma,
formerly trapped by its closed field, to escape.

A jet model with a single magnetic neutral point such as Yokoyama \& Shibata's (1996) anenome jet
model (see their Figure 1) could also result in a plume. Energy gained from emerging flux is
converted to kinetic and thermal energy at the X-type neutral point during reconnection producing a
jet of energized plasma. When the field has reconnected, there is a bundle of newly-opened
magnetic flux through which hitherto trapped coronal plasma can escape as a plume. 

The present results would benefit from future, more extensive analysis of larger data samples
recorded by different instruments in a simultaneous fashion over large time intervals.

\acknowledgements
The authors would like to thank the anonymous referee and J. W. Harvey for helpful comments on the
manuscript. The National Solar Observatory (NSO) is operated by the Association of Universities for
Research in Astronomy, Inc., under cooperative agreement with the National Science Foundation.
SOLIS data used here are produced cooperatively by NSF/NSO and NASA/LWS. NER's work is supported by
NASA grant NNH05AA12I. NER is a member of the coronal polar plume study team sponsored by the
International Space Science Institute (ISSI), Bern, Switzerland.

\clearpage

%% The following command ends your manuscript. LaTeX will ignore any text

\end{document}